\documentclass[lettersize,onecolumn,draft, 12pt]{IEEEtran}
\usepackage{amsmath,amsfonts}
\usepackage{amssymb}
\usepackage{bm}
\usepackage{algorithm}
\usepackage{array}
\usepackage[caption=false,font=normalsize,labelfont=sf,textfont=sf]{subfig}
\usepackage{textcomp}
\usepackage{stfloats}
\usepackage{url}
\usepackage{verbatim}
\usepackage{graphicx}
\usepackage{cite}
\usepackage{mathtools}
\usepackage{color}
\usepackage{algpseudocode}
\usepackage{ifthen}

\newtheorem{Theorem}{Theorem}

\newtheorem{Definition}{Definition}
\newtheorem{Lemma}{Lemma}

\newtheorem{Remark}{Remark}

\begin{document}

\title{Soft Guessing Under Log-Loss Distortion Allowing Errors}

\author{Shota~Saito
\thanks{This work was supported in part by JSPS KAKENHI Grant Numbers JP22K14254, JP23K11097, and JP23H00468.}
\thanks{Shota Saito is with the Faculty of Informatics, Center for Mathematics and Data Science, Gunma University, 4-2, Maebashi, Gunma 371-8510, JAPAN (e-mail: shota.s@gunma-u.ac.jp)}
}


\IEEEpubid{0000--0000/00\$00.00~\copyright~2021 IEEE}

\maketitle

\begin{abstract}
This paper deals with the problem of soft guessing under log-loss distortion (logarithmic loss) that was recently investigated by [Wu and Joudeh, IEEE ISIT, pp. 466--471, 2023]. We extend this problem to soft guessing allowing errors, i.e., at each step, a guesser decides whether to stop the guess or not with some probability and if the guesser stops guessing, then the guesser declares an error. We show that the minimal expected value of the cost of guessing under the constraint of the error probability is characterized by the smooth R\'enyi entropy. Furthermore, we carry out an asymptotic analysis for a stationary and memoryless source.
\end{abstract}


\section{Introduction}\label{introduction}
In information-theoretic literature, the problem of guessing is one of the research topics. In 1994, Massey \cite{Massey} pioneered the information-theoretic study on the problem of guessing and showed that the average number of guesses is characterized by the Shannon entropy. Two years later, Arikan \cite{Arikan} proved that the guessing moment is characterized by the R\'enyi entropy \cite{Renyi}. Since then, the problem of guessing has been studied in various contexts such as guessing subject to distortion \cite{ArikanMerhav}, \cite{Cohen}, \cite{Merhav1999}, \cite{Saito}, \cite{Wu}, guessing allowing errors \cite{Kuzuoka}, \cite{Sakai}, guessing under source uncertainty \cite{Sundaresan}, a large deviation approach \cite{Christiansen}, \cite{Hanawal}, \cite{SundaresanISIT}, joint source-channel coding and guessing \cite{ArikanMerhav2}, guessing via an unreliable oracle \cite{Burin}, guessing with limited memory \cite{Huleihel}, guessing for Markov sources \cite{Malone} and for stationary measures \cite{Pfister}, multi-agent guesswork \cite{Salamatian}, \cite{Salamatian2}, guesswork of hash functions \cite{Yona}, multi-user guesswork \cite{Christiansen2}, guesswork subject to a per-symbol Shannon entropy budget \cite{Beirami}, universal randomized guessing \cite{Cohen}, \cite{Merhav}, guessing based on compressed side information \cite{Graczyk}, guessing individual sequences \cite{Merhav2020}, guesses transmitted via a noisy channel \cite{MerhavNoisyGuess}, multiple guesses under a tunable loss function \cite{Kurri}, and so on.

Recently, Wu and Joudeh \cite{Wu} have investigated the problem of soft guessing. This problem can be seen as a variant of the guessing subject to distortion, but instead of finding a ``hard'' reconstruction (a reproduction symbol) $\hat{x}$, a guesser is interested in a ``soft'' reconstruction (a probability distribution) $\hat{P}$. As a distortion measure, log-loss distortion (logarithmic loss) was adopted in \cite{Wu}. It should be noted that log-loss distortion is widely used in lossy source coding problems (see, e.g., \cite{Courtade}, \cite{Shkel}). For this problem, Wu and Joudeh \cite{Wu} proved that the minimal $\rho$-th guessing moment is characterized by using the R\'enyi entropy of order $1/(1+\rho)$.

In this paper, we extend the problem of soft guessing \cite{Wu} by considering error probability. Specifically, we adopt the framework of guessing allowing errors proposed by Kuzuoka \cite{Kuzuoka}. In the setup of \cite{Kuzuoka}, at the $i$-th step, a guesser decides whether to stop the guess or not; with probability $\pi_i$ ($0 \leq \pi_i \leq 1$), the guesser stops guessing and declares an error; with probability $1-\pi_i$, the guesser continues guessing. For the problem of soft guessing allowing errors, we show that the minimal $\rho$-th guessing moment\footnote{More precisely, we consider a ``cost'' of guessing. For details, please refer to Section \ref{Main}.} under the constraint that the error probability is smaller than $\epsilon$ is characterized by $\epsilon$-smooth R\'enyi entropy \cite{Renner05} of order $1/(1+\rho)$. Furthermore, we investigate an asymptotic formula of the minimal guessing moment for a stationary and memoryless source.

The rest of the paper is organized as follows. In Section \ref{Pre}, we define basic notations, provide the definition of the $\epsilon$-smooth R\'enyi entropy of order $\alpha$ proposed by Renner and Wolf \cite{Renner05}, and show some properties of the smooth R\'enyi entropy. In Section \ref{Main}, we explain the setup of soft guessing allowing errors and we give one-shot upper and lower bounds of the fundamental limit, which are proved in Section \ref{Proof}. In Section \ref{Asymptotic}, we carry out an asymptotic analysis. Finally, Section \ref{Conclusion} is a concluding remark.

\section{Preliminaries} \label{Pre}
\subsection{Basic Notations}
Random variables are denoted by uppercase letters such as $X$, $Y$, and $Z$, and the realizations of random variables are denoted by lowercase letters such as $x$, $y$, and $z$. A set in which a random variable takes values is denoted by a calligraphic font. For example, a random variable $X$ takes values in a set $\mathcal{X}$. The $n$-fold Cartesian product of $\mathcal{X}$ is denoted by $\mathcal{X}^n$. In this paper, we assume that all random variables take values in a finite set. A probability distribution of $X$ is denoted by $P_X$ and a set of probability distributions on $\mathcal{X}$ is denoted by $\mathcal{P}(\mathcal{X})$. An expectation with respect to $P_X$ is denoted by $\mathbb{E}_{P_X}$. We use conventional notations of a joint probability distribution and a conditional probability distribution, e.g., $P_{X,Y}$ and $P_{X|Y}$ denote a joint probability distribution of $X, Y$ and a conditional probability distribution of $X$ given $Y$, respectively. Cardinality of a set is denoted by $|\cdot|$. For $a \in \mathbb{R}$, $\lfloor a \rfloor$ is the greatest integer less than or equal to $a$ and $\lceil a \rceil$ is the least integer greater than or equal to $a$. Throughout this paper, $\log (\cdot) = \log_2 (\cdot)$ and $\exp(\cdot) = 2^{(\cdot)}$.

\subsection{Smooth R\'enyi Entropy} \label{Definition_SRE}
The notion of the $\epsilon$-smooth R\'enyi entropy of order $\alpha$ for $\epsilon \in [0,1)$ and $\alpha \in (0,1) \cup (1,\infty)$, which has been introduced in Renner and Wolf \cite{Renner05}, is defined as
\begin{align}
H^{\epsilon}_{\alpha}(X) := \frac{1}{1 - \alpha} \log \left ( \inf_{Q_X \in \mathcal{B}^{\epsilon}(P_X) } \sum_{x \in \mathcal{X}} [Q_X(x)]^{\alpha} \right ), \label{SmoothRenyiEntropy}
\end{align}
where $\mathcal{B}^{\epsilon}(P_X)$ is a set of functions $Q_X: \mathcal{X} \rightarrow [0,1]$ such that $Q_X(x) \leq P_X(x)$ for all $x \in \mathcal{X}$ and $\sum_{x \in \mathcal{X}} Q_X(x) \geq 1 -\epsilon$.
From this definition, it is clear that $H^{0}_{\alpha}(X)$ (i.e., $H^{\epsilon}_{\alpha}(X)$ for $\epsilon=0$) is equal to the R\'enyi entropy of order $\alpha$ \cite{Renyi}, denoted by $H_{\alpha}(X)$.
For random variables $X$ and $Y$, $H^{\epsilon}_{\alpha}(X,Y)$---a joint version of the $\epsilon$-smooth R\'enyi entropy of order $\alpha$---is defined in a similar manner as \eqref{SmoothRenyiEntropy}.

Furthermore, $H^{\epsilon}_{\alpha}(X|Y)$---the conditional $\epsilon$-smooth R\'enyi entropy of order $\alpha$ of $X$ given $Y$ for $\epsilon \in [0,1)$ and $\alpha \in (0,1) \cup (1,\infty)$---has also proposed in \cite{Renner05} as 
\begin{align}
H^{\epsilon}_{\alpha}(X|Y) := \frac{1}{1 - \alpha} \log \left ( \inf_{Q_{X,Y} \in \mathcal{B}^{\epsilon}(P_{X,Y}) } \max_{y \in \mathcal{Y}} \sum_{x \in \mathcal{X}} \left[\frac{Q_{X,Y}(x,y)}{P_Y(y)} \right]^{\alpha} \right ), \label{Definition_ConditionalSRE}
\end{align}
where $\mathcal{B}^{\epsilon}(P_{X,Y})$ is a set of functions $Q_{X,Y}: \mathcal{X} \times \mathcal{Y} \rightarrow [0,1]$ such that $Q_{X,Y}(x,y) \leq P_{X,Y}(x,y)$ for all $x \in \mathcal{X}$, $y \in \mathcal{Y}$, and $\sum_{x \in \mathcal{X}, y \in \mathcal{Y}} Q_{X,Y}(x,y) \geq 1 -\epsilon$.
In particular, setting $\epsilon=0$ in \eqref{Definition_ConditionalSRE}, it holds that
\begin{align}
H^{0}_{\alpha}(X|Y) 
&= \max_{y \in \mathcal{Y}} \frac{1}{1 - \alpha} \log \sum_{x \in \mathcal{X}} \left[\frac{P_{X,Y}(x,y)}{P_Y(y)} \right]^{\alpha} \\
&= \max_{y \in \mathcal{Y}} \frac{1}{1 - \alpha} \log  \sum_{x \in \mathcal{X}} \left[P_{X|Y}(x|y) \right]^{\alpha}. \label{Definition_ConditionalSRE_epsilon=0}
\end{align}

\subsection{Some Properties of Smooth R\'enyi Entropy} \label{Properties_SRE}
\subsubsection{Chain Rule}
As shown in \cite{Renner05}, the smooth R\'enyi entropy has properties like the Shannon entropy. We describe two of them, which will be used in the proof of the converse part of our main result in Section \ref{ProofConverse}.

\begin{Lemma} \cite[Lemma 5]{Renner05} \label{Lemma1}
    Let $\gamma \geq 0$, $\gamma' \geq 0$, and $\alpha \in (0,1)$. Then, 
    \begin{align}
        H^{\gamma + \gamma'}_{\alpha}(X, Y) \leq H^{\gamma}_{\alpha}(X|Y) + H^{\gamma'}_{\alpha}(Y).
    \end{align}
\end{Lemma}

\begin{Lemma} \cite[Lemma 7]{Renner05} \label{Lemma2}
    For $\epsilon \in [0,1)$ and $\alpha \in (0,1) \cup (1,\infty)$, we have
    \begin{align}
        H^{\epsilon}_{\alpha}(X) \leq H^{\epsilon}_{\alpha}(X, Y).
    \end{align}
\end{Lemma}

\subsubsection{Explicit Formula}
As shown in \cite{Koga}, the smooth R\'enyi entropy has the following explicit formula. This formula will be used in the proof of the achievability part of our main result in Section \ref{ProofAchievability}.

\begin{Lemma} \cite[Theorem 1]{Koga} \label{Lemma3}
    Without loss of generality, we assume that $P_X(1) \geq P_X(2) \geq \ldots \geq P_X(|\mathcal{X}|) >0$. For a given $\epsilon \in [0,1)$, let $i^*$ be the minimum integer such that
    \begin{align}
        \sum_{j=1}^{i^*} P_X(j) \geq 1 - \epsilon.
    \end{align}
    Furthermore, by using $i^*$, we define $Q_\epsilon^*(j)$ as
    \begin{align}
        Q_\epsilon^*(j)=
        \begin{cases}
            P_X(j), & j=1, 2, \ldots, i^* - 1, \\
            1 - \epsilon - \sum_{i=1}^{i^* - 1} P_X(i), & j=i^*, \\
            0, & j=i^* + 1, \ldots, |\mathcal{X}|.
        \end{cases}
    \end{align}
    Then, for $\epsilon \in [0,1)$ and $\alpha \in (0,1)$, the smooth R\'enyi entropy is expressed as
    \begin{align}
        H^{\epsilon}_{\alpha}(X) = \frac{1}{1 - \alpha} \log \left ( \sum_{j=1}^{i^*} \left[Q_\epsilon^*(j) \right]^{\alpha} \right ).
    \end{align}
\end{Lemma}

\subsubsection{Asymptotic Formula}
We define the following three information quantities: 
\begin{align}
    H(X) &:= \mathbb{E}_{P_X} \left [ \log \frac{1}{P_X (X)} \right], \\
    V(X) &:= \mathbb{E}_{P_X} \left [ \left( \log \frac{1}{P_X (X)} - H(X) \right)^2 \right], \\
    T(X) &:= \mathbb{E}_{P_X} \left [ \left | \log \frac{1}{P_X (X)} - H(X) \right |^3 \right].
\end{align}
Let $X^n = (X_1, X_2, \ldots, X_n)$ be $n$ independent copies of $X$. Sakai and Tan \cite{Sakai} have proved the following asymptotic expansion of the smooth R\'enyi entropy $H^{\epsilon}_{\alpha}(X^n)$ up to the third-order term. This formula plays a crucial role in the asymptotic analysis in Section \ref{Asymptotic}.

\begin{Lemma} \cite[Theorem 1]{Sakai} \label{Lemma4}
    If $\alpha \in (0,1)$, $\epsilon \in (0,1)$, $V(X) > 0$, and $T(X) < \infty$, then we have
    \begin{align}
        H^{\epsilon}_{\alpha}(X^n) = n H(X) - \sqrt{n V(X)} \Phi^{-1}(\epsilon) - \frac{1}{2(1-\alpha)} \log n + O(1)
    \end{align}
    as $n \to \infty$, where $\Phi^{-1}: (0,1) \to \mathbb{R}$ is the inverse of the Gaussian cumulative distribution function
    \begin{align}
        \Phi(u) = \int_{-\infty}^{u} \frac{1}{\sqrt{2 \pi}} e^{-\frac{t^2}{2}} \mathrm{d}t.
    \end{align}
\end{Lemma}

\section{Soft Guessing Allowing Errors} \label{Main}
\subsection{Problem Setup} \label{Setup}
In what follows, we assume that $\mathcal{X}=\{1, 2, \ldots, M \}$ and $P_X(1) \geq P_X(2) \geq \ldots \geq P_X(M) > 0$ without loss of generality. A guessing strategy $\mathcal{G}$ is defined by $\mathcal{G}=(\hat{P}, \pi)$, where for some integer $N$,
\begin{align}
    &\hat{P}=(\hat{P}_1, \hat{P}_2, \ldots, \hat{P}_N), \\
    &\hat{P}_i \in \mathcal{P}(\mathcal{X}), \quad i=1, 2, \ldots, N,
\end{align}
and
\begin{align}
    &\pi=(\pi_1, \pi_2, \ldots, \pi_N), \\
    &0 \leq \pi_i \leq 1, \quad i=1, 2, \ldots, N.
\end{align}
When $X=x$, a guesser corresponding to the guessing strategy $\mathcal{G}$ seeks to find a soft reconstruction of $x$ as follows. At the $j$-th step ($j=1, 2, \ldots, N$), the guesser decides whether to stop the guess or not; with probability $\pi_j$, the guesser stops guessing and declares an error; with probability $1-\pi_j$, the guesser does not give up and asks ``Is $d(x, \hat{P}_j) \leq D$?'', where $d(x, \hat{P}_j)$ is the log-loss distortion defined by 
\begin{align}
    d(x, \hat{P}_j) := \log \frac{1}{\hat{P}_j (x)} 
\end{align}
and $D \geq 0$ is a predetermined distortion level. 

In this paper, we consider a $D$-admissible guessing strategy defined as follows (the notion of the $D$-admissible guessing strategy was introduced in \cite{ArikanMerhav}):
\begin{Definition}
    If $\mathbb{P}[d(X, \hat{P}_j) \leq D~{\rm for~some~} j]=1$, then the guessing strategy is called $D$-admissible. A $D$-admissible guessing strategy is denoted by $\mathcal{G}(D)$.
\end{Definition}

\begin{Remark}
    For any $D \geq 0$, an obvious $D$-admissible guessing strategy is given as follows: Let $N=M$ and let $(\hat{P}, \pi)$ be
\begin{align}
    &\hat{P}=(\hat{P}_1, \hat{P}_2, \ldots, \hat{P}_M), \\
    &\hat{P}_j = (\hat{P}_j (1), \hat{P}_j (2), \ldots, \hat{P}_j (M)) \in \mathcal{P}(\mathcal{X}), \\
    &\hat{P}_j (k) =
    \begin{cases}
        1, & k = j, \\
        0, & {\rm otherwise},
    \end{cases}\\
    &\pi=(\pi_1, \pi_2, \ldots, \pi_M)=(0, 0, \ldots, 0).
\end{align}
\end{Remark}

For a $D$-admissible guessing strategy $\mathcal{G}(D)$, the guessing continues until 
(i) when the soft reconstruction of $x$ is found (i.e., $d(x, \hat{P}_j) \leq D$) at the $j$-th step ($j=1, 2, \ldots, N$) 
or
(ii) when the guesser stops guessing and declares an error.
In case (i), the guessing function, which is denoted by $G(x)$, is defined as $G(x)=j$. In other words, the guessing function $G(x)$ induced by the $D$-admissible guessing strategy $\mathcal{G}(D)$ is the minimum index $j \in \{1, 2, \ldots, N \}$ for which $d(x, \hat{P}_j) \leq D$. 
In case (ii), the guessing function is defined as $G(x)=0$ (this definition is used in the previous study \cite{Sakai}).

Given a $D$-admissible guessing strategy $\mathcal{G}(D)$, let 
\begin{align}
    \lambda_i := \prod_{j=1}^{i} (1-\pi_j), \quad i=1, 2, \ldots, N,
\end{align}
and let $Z$ be a random variable taking values in $\{1, 2, \ldots, N\}$ defined by
\begin{align}
    Z := G(X). \label{Definition_Z}
\end{align}
For $z \in \{1, 2, \ldots, N\}$, the probability mass function of $Z$ is given by
\begin{align}
    P_Z(z) = \sum_{x \in G^{-1}(z)} P_X(x),
\end{align}
where $G^{-1}$ is the preimage of $G$, i.e., $G^{-1}(z) := \{x \in \mathcal{X} : G(x) = z \}$.

Given a $D$-admissible guessing strategy $\mathcal{G}(D)$, the error probability and the expected value of the cost are defined as follows:

\begin{Definition}
    The probability of the event that the soft reconstruction of $X$ is found at the $i$-th step before stopping the guess is 
    \begin{align}
        \lambda_i P_Z(i).
    \end{align}
    Hence, the error probability $P_e (G)$, i.e., the probability of the event that the guesser stops guessing and declares an error before finding the soft reconstruction of $X$, is 
    \begin{align}
        P_e (G) = 1 - \sum_{i=1}^{N} \lambda_i P_Z(i). \label{ErrorProbability}
    \end{align}
\end{Definition}

\begin{Definition}
    If the guessing is stopped and an error is declared, then a constant cost $C_e \geq 0$ is incurred as a penalty. On the other hand, if $d(x, \hat{P}_i) \leq D$ at the $i$-th step, the cost of guessing is given by $i^\rho$, where $\rho > 0$ is a constant. Thus, the expected value of the cost is defined as
    \begin{align}
        \sum_{i=1}^{N} \lambda_i P_Z(i) i^{\rho} + P_e (G) C_e. \label{TotalExpectedCost}
    \end{align}
    As in \cite{Kuzuoka}, we only consider the first term in \eqref{TotalExpectedCost}\footnote{The reason why we only consider the first term in \eqref{TotalExpectedCost} is described on page 1678 in \cite{Kuzuoka}.} and it is denoted by $\overline{C}_{\rho, D}(G)$, i.e., 
    \begin{align}
        \overline{C}_{\rho, D}(G) := \sum_{i=1}^{N} \lambda_i P_Z(i) i^{\rho}. \label{ExpectedCost}
    \end{align}
\end{Definition}

\begin{Remark} \label{ConnectionPrevious}
    Special cases of the above setup reduce to the setups of the previous studies \cite{Kuzuoka} and \cite{Wu}. Specifically, the guessing problem considered in \cite{Wu} corresponds to the case where $P_e (G) = 0$. The guessing problem considered in \cite{Kuzuoka} corresponds to the case where $D=0$.
\end{Remark}

\subsection{One-Shot Upper and Lower Bounds of Minimal Expected Value of Cost} \label{MainResults}
We consider the minimization problem of $\overline{C}_{\rho, D}(G)$ under the constraint $P_e (G) \leq \epsilon$, where $\epsilon \in [0, 1)$ and the minimization is over all $D$-admissible guessing strategies. In other words, the fundamental limit we investigate in this paper is 
\begin{align}
    C^\star_X (D, \rho, \epsilon) := \min_{\mathcal{G}(D): P_e (G) \leq \epsilon} \overline{C}_{\rho, D}(G). 
\end{align}
Since we are interested in $C^\star_X (D, \rho, \epsilon)$, we assume $N \leq M$ throughout the paper. This assumption is also imposed in \cite{Wu} (see page 468 in \cite{Wu} in detail).

The lower and upper bounds of $C^\star_X (D, \rho, \epsilon)$ are given as follows:
\begin{Theorem} \label{MainTheorem_OneShot}
    For any $D \geq 0$, $\rho > 0$, and $\epsilon \in [0,1)$, we have
    \begin{align}
        C^\star_X (D, \rho, \epsilon) \geq (1+\log M)^{-\rho} \exp \left \{ \rho H^{\epsilon}_{\frac{1}{1+\rho}}(X) - (1+\rho) \log \lfloor \exp(D) \rfloor \right \}, \label{Converse}
    \end{align}
    and
    \begin{align}
        C^\star_X (D, \rho, \epsilon) \leq 1-\epsilon + 2^\rho \exp \left \{ \rho H^{\epsilon}_{\frac{1}{1+\rho}}(X) - \rho \log \lfloor \exp(D) \rfloor \right \}. \label{Achievability}
    \end{align}
\end{Theorem}

\begin{IEEEproof}
    The lower bound \eqref{Converse} (i.e., the converse part) is proved in Section \ref{ProofConverse} and the upper bound \eqref{Achievability} (i.e., the achievability part) is proved in Section \ref{ProofAchievability}.
\end{IEEEproof}

\begin{Remark}
    When $\epsilon=0$, we have $P_e (G) = 0$. Then, as we have described in Remark \ref{ConnectionPrevious}, the problem reduces to the problem considered in \cite{Wu}. Setting $\epsilon=0$ in Theorem \ref{MainTheorem_OneShot} and recalling that $H^{0}_{\alpha}(X)$ is equal to the R\'enyi entropy $H_{\alpha}(X)$ (see Section \ref{Definition_SRE}), we have
    \begin{align}
        C^\star_X (D, \rho, 0) \geq (1+\log M)^{-\rho} \exp \left \{ \rho H_{\frac{1}{1+\rho}}(X) - (1+\rho) \log \lfloor \exp(D) \rfloor \right \}, \label{ConverseSpecial}
    \end{align}
    and
    \begin{align}
        C^\star_X (D, \rho, 0) \leq 1 + 2^\rho \exp \left \{ \rho H_{\frac{1}{1+\rho}}(X) - \rho \log \lfloor \exp(D) \rfloor \right \}. \label{AchievabilitySpecial}
    \end{align}
    Comparing \eqref{ConverseSpecial} and \eqref{AchievabilitySpecial} to Theorem 1 in \cite{Wu}, we see that the lower bound \eqref{ConverseSpecial} is weaker than that of Theorem 1 in \cite{Wu} and the upper bound \eqref{AchievabilitySpecial} is the same as that of Theorem 1 in \cite{Wu}.
\end{Remark}

\begin{Remark}
    Next, we consider the case $D=0$. Then, as we have described in Remark \ref{ConnectionPrevious}, the problem reduces to the problem considered in \cite{Kuzuoka}. Setting $D=0$ in Theorem \ref{MainTheorem_OneShot}, we have
    \begin{align}
        C^\star_X (0, \rho, \epsilon) \geq (1+\log M)^{-\rho} \exp \left \{ \rho H^{\epsilon}_{\frac{1}{1+\rho}}(X) \right \}, \label{ConverseSpecial2}
    \end{align}
    and
    \begin{align}
        C^\star_X (0, \rho, \epsilon) \leq 1-\epsilon + 2^\rho \exp \left \{ \rho H^{\epsilon}_{\frac{1}{1+\rho}}(X) \right \}. \label{AchievabilitySpecial2}
    \end{align}
    Comparing \eqref{ConverseSpecial2} and \eqref{AchievabilitySpecial2} to Theorems 3 and 4 in \cite{Kuzuoka}, we see that the lower bound \eqref{ConverseSpecial2} is the same as that of Theorem 3 in \cite{Kuzuoka} and the upper bound \eqref{AchievabilitySpecial2} is weaker than that of Theorem 4 in \cite{Kuzuoka}.
\end{Remark}

\section{Proof of Theorem \ref{MainTheorem_OneShot}} \label{Proof}
\subsection{Proof of \eqref{Converse}} \label{ProofConverse}
Given an arbitrary $D$-admissible guessing strategy $\mathcal{G}(D)$ satisfying $P_e (G) \leq \epsilon$, let $Q(i) := \lambda_i P_Z(i)$.
Then, 
\begin{align}
    \sum_{i=1}^N Q(i) 
    &= \sum_{i=1}^N \lambda_i P_Z(i) \\
    &\geq 1 - \epsilon,
\end{align}
where the final inequality is due to \eqref{ErrorProbability} and the assumption $P_e (G) \leq \epsilon$.
Moreover, since $0 \leq \lambda_i \leq 1$ for $i=1, 2, \ldots, N$, we have
\begin{align}
    Q(i) \leq P_Z(i), \quad i=1, 2, \ldots, N.
\end{align}
Hence, it holds that
\begin{align}
    Q \in \mathcal{B}^{\epsilon}(P_Z). \label{Condition_Q}
\end{align}

We use the following lemma introduced in \cite{Arikan}:
\begin{Lemma} \cite[Lemma 1]{Arikan}
    For non-negative numbers $a_i$ and $b_i$ ($i=1, 2, \ldots, N$) and any $\theta \in (0, 1)$, we have
    \begin{align}
        \sum_{i=1}^N a_i b_i \geq \left (\sum_{i=1}^N a_i^{\frac{-\theta}{1-\theta}} \right)^{\frac{1-\theta}{-\theta}} \left (\sum_{i=1}^N b_i^\theta \right )^{\frac{1}{\theta}}. \label{ConverseLemma}
    \end{align}
\end{Lemma}
Setting $a_i = i^\rho$, $b_i = Q(i)$, and $\theta=1/(1+\rho)$ in \eqref{ConverseLemma}, 
\begin{align}
    \overline{C}_{\rho, D}(G)
    =\sum_{i=1}^N Q(i) i^\rho 
    &\geq \left (\sum_{i=1}^N i^{-1} \right)^{-\rho} \left (\sum_{i=1}^N Q(i)^{\frac{1}{1+\rho}} \right )^{1+\rho} \\
    &\geq (1+\log M)^{-\rho} \exp \left \{ \rho H^{\epsilon}_{\frac{1}{1+\rho}}(Z) \right \}, \label{Converse1}
\end{align}
where the last inequality follows from $\sum_{i=1}^N i^{-1} \leq 1 + \log N \leq 1 + \log M$ and \eqref{Condition_Q}.\footnote{As we have explained at the beginning of Section \ref{MainResults}, $N \leq M$ is assumed.} 

To complete the proof, we utilize the properties of the smooth R\'enyi entropy described in Section \ref{Properties_SRE}. Since $\rho > 0$, it holds that $0 < 1/(1+\rho) <1$. Hence, we have
\begin{align}
     H^{\epsilon}_{\frac{1}{1+\rho}}(Z)
     & \overset{(a)}{\geq} H^{\epsilon}_{\frac{1}{1+\rho}}(X, Z) - H^{0}_{\frac{1}{1+\rho}}(X|Z) \\
     & \overset{(b)}{\geq} H^{\epsilon}_{\frac{1}{1+\rho}}(X) - H^{0}_{\frac{1}{1+\rho}}(X|G(X)), \label{Converse2}
\end{align}
where $(a)$ follows from Lemma \ref{Lemma1} and $(b)$ follows from Lemma \ref{Lemma2} and \eqref{Definition_Z}.

Finally, we evaluate $H^{0}_{\frac{1}{1+\rho}}(X|G(X))$. From \eqref{Definition_ConditionalSRE_epsilon=0}, 
\begin{align}
    H^{0}_{\frac{1}{1+\rho}}(X|G(X)) = \max_{z \in \{1, 2, \ldots, N \}} \frac{1+\rho}{\rho} \log \sum_{x \in \mathcal{X}} [P_{X|G(X)} (x|z)]^{\frac{1}{1+\rho}}. \label{ConditionalSREofXgivenG(X)}
\end{align}
Letting $z^*$ be the element of $\{1, 2, \ldots, N \}$ that attains the maximum in \eqref{ConditionalSREofXgivenG(X)}, we have
\begin{align}
    H^{0}_{\frac{1}{1+\rho}}(X|G(X)) = \frac{1+\rho}{\rho} \log \sum_{x \in \mathcal{X}} [P_{X|G(X)} (x|z^*)]^{\frac{1}{1+\rho}} 
\end{align}
and this is upper bounded as 
\begin{align}
    H^{0}_{\frac{1}{1+\rho}}(X|G(X)) 
    & \overset{(a)}{=} \frac{1+\rho}{\rho} \log \sum_{x \in G^{-1}(z^*)} [P_{X|G(X)} (x|z^*)]^{\frac{1}{1+\rho}} \\
    & \leq \frac{1+\rho}{\rho} \log \sum_{x \in G^{-1}(z^*)} 1^{\frac{1}{1+\rho}} \\ 
    & = \frac{1+\rho}{\rho} \log |G^{-1}(z^*)| \\
    & \overset{(b)}{\leq} \frac{1+\rho}{\rho} \log \lfloor \exp(D) \rfloor, \label{Converse3}
\end{align}
where $(a)$ is due to the fact that the support of $X$ is $G^{-1}(z^*)$ given that $G(X)=z^*$ and $(b)$ follows from Lemma 1 in \cite{Wu}.

Combining \eqref{Converse1}, \eqref{Converse2}, and \eqref{Converse3}, we obtain
\begin{align}
    \overline{C}_{\rho, D}(G) \geq (1+\log M)^{-\rho} \exp \left \{\rho H^{\epsilon}_{\frac{1}{1+\rho}}(X) - (1+\rho) \log \lfloor \exp(D) \rfloor \right \},
\end{align}
and since $\mathcal{G}(D)$ is an arbitrary $D$-admissible guessing strategy satisfying $P_e (G) \leq \epsilon$, we have \eqref{Converse} and complete the proof.

\subsection{Proof of \eqref{Achievability}} \label{ProofAchievability}
The key points of the proof are the explicit formula of the $\epsilon$-smooth R\'enyi entropy of order $\alpha$ given in Lemma \ref{Lemma3} and the similar technique used in \cite{Wu}.
As we have described in Section \ref{Setup}, recall that $\mathcal{X}=\{1, 2, \ldots, M \}$ and $P_X(1) \geq P_X(2) \geq \ldots \geq P_X(M) > 0$.

First, let
\begin{align}
    K &:= \left \lceil \frac{i^*}{\lfloor \exp(D) \rfloor} \right \rceil, \\
    K' &:= \left \lceil \frac{M - i^*}{\lfloor \exp(D) \rfloor} \right \rceil, \\
    N &:= K + K',
\end{align}
where $i^*$ is defined in Lemma \ref{Lemma3}. Second, we construct $\hat{P}^\dagger=(\hat{P}_1^\dagger, \hat{P}_2^\dagger, \ldots, \hat{P}_N^\dagger)$ as follows: Let $\mathcal{L}_1, \mathcal{L}_2, \ldots, \mathcal{L}_{K}$ be
\begin{align}
    \mathcal{L}_i &:= \{(i-1) \lfloor \exp(D) \rfloor+1,  (i-1) \lfloor \exp(D) \rfloor +2, \ldots, i \lfloor \exp(D) \rfloor \}
\end{align}
for $i=1, 2, \ldots, K-1$ and
\begin{align}
    \mathcal{L}_K &:= \{(K-1) \lfloor \exp(D) \rfloor+1,  (K-1) \lfloor \exp(D) \rfloor +2, \ldots, i^* \}. 
\end{align}
Moreover, let $\mathcal{L}'_1, \mathcal{L}'_2, \ldots, \mathcal{L}'_{K'}$ be
\begin{align}
    \mathcal{L}'_j &:= \{(j-1) \lfloor \exp(D) \rfloor+i^*+1,  (j-1) \lfloor \exp(D) \rfloor + i^* +2, \ldots, j \lfloor \exp(D) \rfloor +i^* \}
\end{align}
for $j=1, 2, \ldots, K'-1$ and 
\begin{align}
    \mathcal{L}'_{K'} &:= \{(K'-1) \lfloor \exp(D) \rfloor+i^*+1,  (K'-1) \lfloor \exp(D) \rfloor +i^*+2, \ldots, M \}.
\end{align}
From the above construction, it holds that
\begin{align}
    \mathcal{X} = \bigcup_{j=1}^{K} \mathcal{L}_j \cup \bigcup_{j=1}^{K'} \mathcal{L}'_j. \label{Partition}
\end{align}
Then, $\hat{P}_1^\dagger, \hat{P}_2^\dagger, \ldots, \hat{P}_N^\dagger$ are defined as
\begin{align}
    \hat{P}_i^\dagger(x) &:= \frac{1}{\lfloor \exp(D) \rfloor}, \quad \forall x \in \mathcal{L}_i, \quad (i=1, 2, \ldots, K-1), \\
    \hat{P}_K^\dagger(x) &:= \frac{1}{i^* - (K-1) \lfloor \exp(D) \rfloor}, \quad \forall x \in \mathcal{L}_K,\\
    \hat{P}_{K+j}^\dagger(x) &:=\frac{1}{\lfloor \exp(D) \rfloor}, \quad \forall x \in \mathcal{L}'_j, \quad (j=1, 2, \ldots, K'-1), \\
    \hat{P}_{N}^\dagger(x) &:= \frac{1}{M - (K'-1) \lfloor \exp(D) \rfloor-i^*}, \quad \forall x \in \mathcal{L}'_{K'}.
\end{align}
Next, we construct $\pi^\dagger=(\pi_1^\dagger, \pi_2^\dagger, \ldots, \pi_N^\dagger)$ as 
\begin{align}
    \pi_j^\dagger=
    \begin{cases}
        0, & j=1, 2, \ldots, K-1, \\
        1 - \frac{\sum_{i=(K-1) \lfloor \exp(D) \rfloor+1}^{i^*} Q_\epsilon^*(i)}{\sum_{i=(K-1) \lfloor \exp(D) \rfloor+1}^{i^*} P_X (i)}, & j=K, \\
        1, & j=K+1, K+2, \ldots, N,
    \end{cases}
\end{align}
and let
\begin{align}
    \lambda_i^\dagger := \prod_{j=1}^{i} (1-\pi_j^\dagger).
\end{align}
It should be noted that the above guessing strategy $(\hat{P}^\dagger, \pi^\dagger)$ is $D$-admissible because any $x \in \mathcal{X}$ is in $\mathcal{L}_j$ or $\mathcal{L}'_j$ (see \eqref{Partition}) and thus for any $x \in \mathcal{X}$, there exists $\hat{P}_j^\dagger$ such that
\begin{align}
    d(x, \hat{P}_j^\dagger)
    &=\log \frac{1}{\hat{P}_j^\dagger (x)} \\
    &\leq \log \lfloor \exp(D) \rfloor \\
    &\leq D.
\end{align}

Let the guessing function induced by the guessing strategy $(\hat{P}^\dagger, \pi^\dagger)$ be denoted by $G^\dagger$ and $Z^\dagger := G^\dagger (X)$. The expected value of the cost $\overline{C}_{\rho, D}(G^\dagger)$ is calculated as
\begin{align}
    \overline{C}_{\rho, D}(G^\dagger) 
    &= \sum_{i=1}^{N} \lambda_i^\dagger P_{Z^\dagger} (i) i^{\rho} \\
    & \overset{(a)}{=} \sum_{j=1}^{i^*} Q_\epsilon^*(j) \left \lceil \frac{j}{\lfloor \exp(D) \rfloor} \right \rceil^\rho \\
    &= \sum_{j=1}^{i^*} Q_\epsilon^*(j) \left \lceil \frac{1}{\lfloor \exp(D) \rfloor} \sum_{k : k \leq j} 1 \right \rceil^\rho \\
    &\overset{(b)}{\leq} \sum_{j=1}^{i^*} Q_\epsilon^*(j) \left \lceil \frac{1}{\lfloor \exp(D) \rfloor} \sum_{k : k \leq j} \left( \frac{Q_\epsilon^*(k)}{Q_\epsilon^*(j)} \right)^{\frac{1}{1+\rho}} \right \rceil^\rho \\
    &\leq \sum_{j=1}^{i^*} Q_\epsilon^*(j) \left \lceil \frac{1}{\lfloor \exp(D) \rfloor} \sum_{k=1}^{i^*} \left( \frac{Q_\epsilon^*(k)}{Q_\epsilon^*(j)} \right)^{\frac{1}{1+\rho}} \right \rceil^\rho \\
    &\overset{(c)}{\leq} \sum_{j=1}^{i^*} Q_\epsilon^*(j) \left \{1+2^\rho \left( \frac{1}{\lfloor \exp(D) \rfloor} \sum_{k=1}^{i^*} \left( \frac{Q_\epsilon^*(k)}{Q_\epsilon^*(j)} \right)^{\frac{1}{1+\rho}} \right)^\rho \right \} \\
    &= \sum_{j=1}^{i^*} Q_\epsilon^*(j) + \left( \frac{2}{\lfloor \exp(D) \rfloor} \right)^\rho \left (\sum_{j=1}^{i^*} [Q_\epsilon^*(j)]^{\frac{1}{1+\rho}} \right )^{1+\rho} \\
    &\overset{(d)}{=} 1-\epsilon + \left( \frac{2}{\lfloor \exp(D) \rfloor} \right)^\rho \exp \left \{ \rho H^{\epsilon}_{\frac{1}{1+\rho}}(X) \right \} \\
    &= 1-\epsilon + 2^\rho \exp \left \{ \rho H^{\epsilon}_{\frac{1}{1+\rho}}(X) - \rho \log \lfloor \exp(D) \rfloor \right \},
\end{align}
where $(a)$ follows from the definition of the guessing strategy $(\hat{P}^\dagger, \pi^\dagger)$ and the definition of $Q_\epsilon^*$ in Lemma \ref{Lemma3}, $(b)$ is due to the definition of $Q_\epsilon^*$ in Lemma \ref{Lemma3}, $(c)$ comes from the inequality $\lceil \xi \rceil^\rho < 1 + 2^\rho \xi^\rho $ for $\xi \geq 0$ and $\rho > 0$ (see Eq.\ (26) in \cite{Bunte}), and in step $(d)$, we used Lemma \ref{Lemma3}.

Finally, since the guessing strategy $(\hat{P}^\dagger, \pi^\dagger)$ is $D$-admissible and satisfies $P_e (G^\dagger) \leq \epsilon$ (see \eqref{ErrorProbability} and the definition of $(\hat{P}^\dagger, \pi^\dagger)$), we complete the proof of \eqref{Achievability}.

\section{Asymptotic Analysis} \label{Asymptotic}
In this section, we investigate the minimal expected value of the cost for a stationary and memoryless source. Let $X^n = (X_1, X_2, \ldots, X_n)$ be $n$ independent copies of $X$. As in \cite{Shkel} and \cite{Wu}, for $n$-letter setting, the log-loss distortion is defined as
\begin{align}
    d_n (x^n, \hat{P}_i^n) := \frac{1}{n} \log \frac{1}{\hat{P}_i^n (x^n)}, 
\end{align}
where $\hat{P}_i^n \in \mathcal{P}(\mathcal{X}^n)$. The asymptotic expansion of the fundamental limit $C^\star_{X^n} (D, \rho, \epsilon)$ is given as follows:

\begin{Theorem} \label{MainAsymptotic}
    If $D \geq 0$, $\rho > 0$, $\epsilon \in (0,1)$, $V(X) > 0$, and $T(X) < \infty$, then for a stationary and memoryless source, we have
    \begin{align}
        &C^\star_{X^n} (D, \rho, \epsilon) \nonumber \\
        & \quad \geq (1+n \log M)^{-\rho} \nonumber \\
        & \quad \quad \times \exp \left \{\rho n H(X) - \rho \sqrt{n V(X)} \Phi^{-1}(\epsilon) - \frac{1+\rho}{2} \log n - (1+\rho) \log \lfloor \exp(nD) \rfloor + O(1) \right \}, \\
        &C^\star_{X^n} (D, \rho, \epsilon) \nonumber \\
        & \quad \leq 1-\epsilon + 2^\rho \exp \left \{ \rho n H(X) - \rho \sqrt{n V(X)} \Phi^{-1}(\epsilon) - \frac{1+\rho}{2} \log n - \rho \log \lfloor \exp(nD) \rfloor + O(1) \right \}
    \end{align}
    as $n \to \infty$.
\end{Theorem}

\begin{IEEEproof}
    Combining the asymptotic expansion of the smooth R\'enyi entropy given by Lemma \ref{Lemma4} and the one-shot formula given by Theorem \ref{MainTheorem_OneShot} yields Theorem \ref{MainAsymptotic}.
\end{IEEEproof}

\section{Concluding Remark} \label{Conclusion}
We have investigated the problem of soft guessing allowing errors. We have established the one-shot upper and lower bounds of $C^\star_X (D, \rho, \epsilon)$ via the smooth R\'enyi entropy. Moreover, we have provided the asymptotic expansion of $C^\star_{X^n} (D, \rho, \epsilon)$, where $X^n = (X_1, X_2, \ldots, X_n)$ is $n$ independent copies of $X$. 
Future works are as follows:

\subsubsection{Extension}
We can consider extensions of the setup in this paper. One example is the case where side information is available to the guesser as in \cite{Arikan} and \cite{Kuzuoka}. Another example is the case where we adopt a tunable loss (i.e., a generalization of log-loss) as in \cite{Kurri}.

\subsubsection{Connection to lossy source coding}
The cumulant generating function of codeword lengths has been investigated in variable-length lossy source coding (see, e.g., \cite{CourtadeVL}, \cite{SaitoVL}, \cite{Wu}). The author conjectures that the problem considered in this paper is closely related to the variable-length lossy source coding in \cite{SaitoVL}, i.e., the variable-length lossy source coding in which the criteria are the cumulant generating function of codeword lengths and excess distortion probability. 


 

\begin{thebibliography}{99}
\bibitem{Arikan}
E. Arikan, "An inequality on guessing and its application to sequential decoding," in {\it IEEE Transactions on Information Theory}, vol. 42, no. 1, pp. 99--105, Jan. 1996, doi: 10.1109/18.481781.

\bibitem{ArikanMerhav}
E. Arikan and N. Merhav, "Guessing subject to distortion," in {\it IEEE Transactions on Information Theory}, vol. 44, no. 3, pp. 1041--1056, May 1998, doi: 10.1109/18.669158.

\bibitem{ArikanMerhav2}
E. Arikan and N. Merhav, "Joint source-channel coding and guessing with application to sequential decoding," in {\it IEEE Transactions on Information Theory}, vol. 44, no. 5, pp. 1756--1769, Sept. 1998, doi: 10.1109/18.705557.

\bibitem{Beirami}
A. Beirami, R. Calderbank, K. Duffy, and M. Médard, "Quantifying computational security subject to source constraints, guesswork and inscrutability," in {\it Proc.\ 2015 IEEE International Symposium on Information Theory (ISIT)}, Hong Kong, China, 2015, pp. 2757--2761, doi: 10.1109/ISIT.2015.7282958.

\bibitem{Bunte}
C. Bunte and A. Lapidoth, "Encoding tasks and Rényi entropy," in {\it IEEE Transactions on Information Theory}, vol. 60, no. 9, pp. 5065-5076, Sept. 2014, doi: 10.1109/TIT.2014.2329490.

\bibitem{Burin}
A. Burin and O. Shayevitz, "Reducing guesswork via an unreliable oracle," in {\it IEEE Transactions on Information Theory}, vol. 64, no. 11, pp. 6941--6953, Nov. 2018, doi: 10.1109/TIT.2018.2856190.

\bibitem{Christiansen}
M. M. Christiansen and K. R. Duffy, "Guesswork, large deviations, and Shannon entropy," in {\it IEEE Transactions on Information Theory}, vol. 59, no. 2, pp. 796--802, Feb. 2013, doi: 10.1109/TIT.2012.2219036.

\bibitem{Christiansen2}
M. M. Christiansen, K. R. Duffy, F. du Pin Calmon, and M. Médard, "Multi-user guesswork and brute force security," in {\it IEEE Transactions on Information Theory}, vol. 61, no. 12, pp. 6876--6886, Dec. 2015, doi: 10.1109/TIT.2015.2482972.

\bibitem{Cohen}
A. Cohen and N. Merhav, "Universal randomized guessing subject to distortion," in {\it IEEE Transactions on Information Theory}, vol. 68, no. 12, pp. 7714--7734, Dec. 2022, doi: 10.1109/TIT.2022.3194073.

\bibitem{Courtade}
T. A. Courtade and T. Weissman, "Multiterminal source coding under logarithmic loss," in {\it IEEE Transactions on Information Theory}, vol. 60, no. 1, pp. 740--761, Jan. 2014, doi: 10.1109/TIT.2013.2288257.

\bibitem{CourtadeVL}
T. A. Courtade and S. Verdú, "Variable-length lossy compression and channel coding: Non-asymptotic converses via cumulant generating functions," in {\it Proc.\ 2014 IEEE International Symposium on Information Theory (ISIT)}, Honolulu, HI, USA, 2014, pp. 2499--2503, doi: 10.1109/ISIT.2014.6875284.

\bibitem{Graczyk}
R. Graczyk, A. Lapidoth, N. Merhav, and C. Pfister, "Guessing based on compressed side information," in {\it IEEE Transactions on Information Theory}, vol. 68, no. 7, pp. 4244--4256, July 2022, doi: 10.1109/TIT.2022.3158475.

\bibitem{Hanawal}
M. K. Hanawal and R. Sundaresan, "Guessing revisited: A large deviations approach," in {\it IEEE Transactions on Information Theory}, vol. 57, no. 1, pp. 70--78, Jan. 2011, doi: 10.1109/TIT.2010.2090221.

\bibitem{Huleihel}
W. Huleihel, S. Salamatian, and M. Médard, "Guessing with limited memory," in {\it Proc.\ 2017 IEEE International Symposium on Information Theory (ISIT)}, Aachen, Germany, 2017, pp. 2253-2257, doi: 10.1109/ISIT.2017.8006930.

\bibitem{Koga}
H. Koga, "Characterization of the smooth Rényi entropy using majorization," in {\it Proc.\ 2013 IEEE Information Theory Workshop (ITW)}, Seville, Spain, 2013, pp. 1-5, doi: 10.1109/ITW.2013.6691332.

\bibitem{Kurri}
G. R. Kurri, O. Kosut, and L. Sankar, "Evaluating multiple guesses by an adversary via a tunable loss function," in {\it Proc.\ 2021 IEEE International Symposium on Information Theory (ISIT)}, Melbourne, Australia, 2021, pp. 2002-2007, doi: 10.1109/ISIT45174.2021.9517733.

\bibitem{Kuzuoka}
S. Kuzuoka, "On the conditional smooth Rényi entropy and its applications in guessing and source coding," in {\it IEEE Transactions on Information Theory}, vol. 66, no. 3, pp. 1674--1690, March 2020, doi: 10.1109/TIT.2019.2937318.

\bibitem{Malone}
D. Malone and W. G. Sullivan, "Guesswork and entropy," in {\it IEEE Transactions on Information Theory}, vol. 50, no. 3, pp. 525--526, March 2004, doi: 10.1109/TIT.2004.824921.

\bibitem{Massey}
J. L. Massey, "Guessing and entropy," in {\it Proc.\ 1994 IEEE International Symposium on Information Theory (ISIT)}, Trondheim, Norway, 1994, pp. 204, doi: 10.1109/ISIT.1994.394764.

\bibitem{Merhav1999}
N. Merhav, R. M. Roth, and E. Arikan, "Hierarchical guessing with a fidelity criterion," in {\it IEEE Transactions on Information Theory}, vol. 45, no. 1, pp. 330--337, Jan. 1999, doi: 10.1109/18.746836.

\bibitem{Merhav}
N. Merhav and A. Cohen, "Universal randomized guessing With application to asynchronous decentralized brute–force attacks," in {\it IEEE Transactions on Information Theory}, vol. 66, no. 1, pp. 114--129, Jan. 2020, doi: 10.1109/TIT.2019.2920538.

\bibitem{Merhav2020}
N. Merhav, "Guessing individual sequences: Generating randomized guesses using finite–state machines," in {\it IEEE Transactions on Information Theory}, vol. 66, no. 5, pp. 2912--2920, May 2020, doi: 10.1109/TIT.2019.2946303.

\bibitem{MerhavNoisyGuess}
N. Merhav, "Noisy guesses," in {\it IEEE Transactions on Information Theory}, vol. 66, no. 8, pp. 4796--4803, Aug. 2020, doi: 10.1109/TIT.2020.2974845.

\bibitem{Pfister}
C. E. Pfister and W. G. Sullivan, "Renyi entropy, guesswork moments, and large deviations," in {\it IEEE Transactions on Information Theory}, vol. 50, no. 11, pp. 2794--2800, Nov. 2004, doi: 10.1109/TIT.2004.836665.

\bibitem{Renner05}
R. Renner and S. Wolf, "Simple and tight bounds for information reconciliation and privacy amplification," in {\it Advances in Cryptology - ASIACRYPT 2005}, Lecture Notes in Computer Science, vol. 3788. Springer, Berlin, Heidelberg. 

\bibitem{Shkel}
Y. Y. Shkel and S. Verdú, "A single-shot approach to lossy source coding under logarithmic loss," in {\it IEEE Transactions on Information Theory}, vol. 64, no. 1, pp. 129--147, Jan. 2018, doi: 10.1109/TIT.2017.2700860.

\bibitem{Saito}
S. Saito and T. Matsushima, "Non-asymptotic fundamental limits of guessing subject to distortion," in {\it Proc.\ 2019 IEEE International Symposium on Information Theory (ISIT)}, Paris, France, 2019, pp. 652-656, doi: 10.1109/ISIT.2019.8849434.

\bibitem{SaitoVL}
S. Saito and T. Matsushima, "Non-asymptotic bounds of cumulant generating function of codeword lengths in variable-length lossy compression," in {\it IEEE Transactions on Information Theory}, vol. 69, no. 4, pp. 2113--2119, April 2023, doi: 10.1109/TIT.2022.3229358.

\bibitem{Sakai}
Y. Sakai and V. Y. F. Tan, "On smooth Rényi entropies: A novel information measure, one-shot coding theorems, and asymptotic expansions," in {\it IEEE Transactions on Information Theory}, vol. 68, no. 3, pp. 1496--1531, March 2022, doi: 10.1109/TIT.2021.3132670.

\bibitem{Salamatian}
S. Salamatian, A. Beirami, A. Cohen, and M. Médard, "Centralized vs decentralized multi-agent guesswork," in {\it Proc.\ 2017 IEEE International Symposium on Information Theory (ISIT)}, Aachen, Germany, 2017, pp. 2258--2262, doi: 10.1109/ISIT.2017.8006931.

\bibitem{Salamatian2}
S. Salamatian, W. Huleihel, A. Beirami, A. Cohen, and M. Médard, "Why botnets work: Distributed brute-force attacks need no synchronization," in {\it IEEE Transactions on Information Forensics and Security}, vol. 14, no. 9, pp. 2288--2299, Sept. 2019, doi: 10.1109/TIFS.2019.2895955.

\bibitem{Sundaresan}
R. Sundaresan, "Guessing under source uncertainty," in {\it IEEE Transactions on Information Theory}, vol. 53, no. 1, pp. 269--287, Jan. 2007, doi: 10.1109/TIT.2006.887466.

\bibitem{SundaresanISIT}
R. Sundaresan, "Guessing based on length functions," in {\it Proc.\ 2007 IEEE International Symposium on Information Theory (ISIT)}, Nice, France, 2007, pp. 716--719, doi: 10.1109/ISIT.2007.4557309.

\bibitem{Renyi}
A. R\'enyi, "On measures of entropy and information," {\it 4th Berkley Symposium on Mathematics, Statistics and Probability}, pp. 547--561, 1961.

\bibitem{Wu}
H. Wu and H. Joudeh, "Soft guessing under logarithmic loss,", in {\it Proc.\ 2023 IEEE International Symposium on Information Theory (ISIT)}, Taipei, Taiwan, 2023, pp. 466--471, doi: 10.1109/ISIT54713.2023.10206698.

\bibitem{Yona}
Y. Yona and S. Diggavi, "The effect of bias on the guesswork of hash functions," in {\it Proc.\ 2017 IEEE International Symposium on Information Theory (ISIT)}, Aachen, Germany, 2017, pp. 2248--2252, doi: 10.1109/ISIT.2017.8006929.
\end{thebibliography}
%

\begin{IEEEbiographynophoto}{Shota Saito} (Member, IEEE) received the B.E., M.E., and Ph.D. degrees in
applied mathematics from Waseda University, Tokyo, Japan, in 2013, 2015, and 2018, respectively.

From 2018 to 2021, he was an Assistant Professor at the Department of Applied Mathematics, Waseda University, Tokyo, Japan. Since 2021, he has been an Associate Professor with the Faculty of Informatics, Gunma University, Gunma, Japan. His research interests include information theory and its applications for machine learning and information-theoretic security.

Dr. Saito is a member of the Institute of Electronics, Information and Communication Engineers (IEICE). He was a recipient of the IEEE IT Society Japan Chapter Young Researcher Best Paper Award and the Student Paper
Award from the 2016 International Symposium on Information Theory and Its Applications. He also received the Waseda University Azusa Ono Memorial Award (Academic) in 2016 and the Symposium on Information Theory and
Its Applications Young Researcher Paper Award in 2018.
\end{IEEEbiographynophoto}

\vfill

\end{document}